\documentclass[12pt]{article}

\usepackage{graphicx}
\usepackage{amsmath,amssymb}
\usepackage{amsfonts}

\newcommand{\ve}[1]{\mathbf{#1}}

\begin{document}

\begin{center}
{\bf ROLE OF VACUUM POLARIZATION FOR THE ANNIHILATION CHANNEL IN A STRONG 
LASER FIELD} 
\vspace*{6mm}

S.~A.~Smolyansky$^{1 \dag}$, D.~B.~Blaschke$^{2,3}$, A.~V.~Chertilin$^{1}$,\\
G.~R\"{o}pke$^{4}$, 
A.~V.~Tarakanov$^{1}$ 

$^1${\small {\it Physics Department of Saratov State University,
410026, Saratov, Russia}}\\
$^2${\small {\it Institute for Theoretical Physics, University of Wroc{\l}aw, 
50-204 Wroc{\l}aw, Poland}} \\
$^3${\small {\it Bogoliubov Laboratory for Theoretical Physics, JINR Dubna, 
141980 Dubna, Russia}}\\
$^4${\small {\it Institut f\"{u}r Physik, Universit\"{a}t Rostock, 
D-18051 Rostock, Germany}}\\
$\dag$ {\small \it
E-mail: smol@sgu.ru
}             
\end{center}
%
\vskip 5mm

\begin{abstract}

We consider vacuum polarization effects in the one-photon annihilation channel
within a kinetic description of the $e^{-} e^{+}$ plasma produced from the 
vacuum in the focal spot of counter-propagating laser beams. 
This entails essential changes in the structure of the photon kinetic equation.
We investigate the domain  of large adiabaticity  parameters $\gamma \gg 1$ 
where the photon radiation turns out to be very small. 
A more thorough examination of the domain $\gamma \lesssim 1$ needs separate 
investigation. 
However, an exploratory study has shown that the one-photon annihilation 
channel can lead for some domains of laser field parameters 
(e.g., for the XFEL) to contributions accessible for observation.

\end{abstract}


\section{Introduction}

The planned experiments \cite{bib1} for the observation of an $e^{-} e^{+}$ 
plasma created from the vacuum in the focal spot of two counter-propagating 
optical laser beams with the intensity $I \gtrsim 10^{21}$ W/cm$^{2}$ raises 
the problem of an accurate theoretical description of the experimental 
manifestations of the dynamical Schwinger effect \cite{bib2},
see also Refs. \cite{Blaschke:2008wf,Hebenstreit:2008ae,Yaresko:2010xe}. 
The existing prediction \cite{bib3} in the domain of strongly sub-critical 
fields $E\ll E_{c}=m^{2}/e$ of a considerable number of secondary annihilation 
photons is not rather convincing because it is based on the S-matrix approach 
for the description of quasiparticle excitations in the presence of a strong 
external electric field. 
In particular, this approach does not take into account vacuum polarization
effects. 
Apparently, an adequate approach for description of vacuum excitations in 
strong electromagnetic fields is a kinetic theory in the quasiparticle 
representation. 
The simplest kinetic equation (KE) of such type for the  $e^{-}$, $e^{+}$ 
subsystem has been obtained for the case of linearly polarized, time dependent 
and spatially homogeneous electric fields \cite{bib2}. 
Some generalizations of the KE in the fermion sector have been worked out in 
Refs.~ \cite{bib4}.

It can be expected, that electromagnetic field fluctuations of the $e^{-}e^{+}$
plasma are accompanied by the generation of real photons which can be 
registered far from the focal spot. 
The first two equations of the BBGKY chain for the photon sector of the  
$e^{-} e^{+} \gamma$ plasma were obtained in \cite{bib6}. 
This level is sufficient for the kinetic description of the one-photon 
annihilation. 
In the presence of an external field such process is not forbidden \cite{bib8}.
In the works \cite{bib6} it was shown that the spectrum of the secondary 
photons in the low frequency domain $k\ll m$ has the character of the flicker 
noise. 
In the present work the inclusion of vacuum polarization effects in the 
one-photon radiation spectrum leading to an essential change of the photon KE 
structure will be investigated in the broad spectral band including the 
annihilation domain $\nu\sim2m$.

We have first considered the domain of large adiabaticity parameters 
$\gamma\gg 1$, where the photon radiation from the focal spot turns out to be 
very small. 
However, the tendency of the effect to grow for $\gamma\rightarrow 1$ has been
discovered. 
This is just the domain of practical interest for parameters of modern lasers. 
We intend to investigate this domain in a next step.

\section{Kinetics of the one-photon annihilation channel}

We construct the photon kinetics as a perturbation theory with respect to the
electron-photon coupling constant while the interaction with an external 
quasi-classical electric field is taken into account exactly by using the 
non-stationary spinor basis \cite{bib4}. 
For a description of the one-photon annihilation process it is sufficient to 
consider the first two equations of the BBGKY chain \cite{bib6}. 
The first equation has the form
\begin{eqnarray}\label{1e}
 \dot{F}_{rr'}(\ve{k},\ve{k}',t) &=&
ie(2\pi)^{-3/2}\int d^3 p_1 d^3 p_2\Bigl\{\frac{1}{\sqrt{2k}} 
\delta(\ve{p}_1 -\ve{p}_2 -\ve{k}) \nonumber  \\ 
& &\times
   [\bar{u}v]^{r}_{\beta\alpha}(\ve{p}_1 ,\ve{p}_2 ,\ve{k};t)
\langle b^{+}_{\beta}(-\ve{p}_{2},t)a^+ _{\alpha}(\ve{p}_1 ,t)
A_{r'} ^{(-)} (\ve{k}',t)\rangle + \frac{1}{\sqrt{2k'}} 
\delta(\ve{p}_1 -\ve{p}_2 +\ve{k}')\nonumber  \\ 
&& \times
  [\bar{v}u]^{r'}_{\beta\alpha}(\ve{p}_1 ,\ve{p}_2 ,\ve{k}';t)
\langle b_{\alpha}(-\ve{p}_{1},t)a_{\beta}(\ve{p}_2 ,t)A_{r} ^{(+)} (\ve{k},t)
\rangle \Bigr\}~,
\end{eqnarray}
where 
$F_{rr'}(\ve{k},\ve{k}',t)= 
\langle A^{(+)}_{r}(\ve{k},t)A^{(-)}_{r'}(\ve{k}',t)\rangle$
is the two-time photon correlation function.

The Heisenberg-like fermion equations of motion contain terms stipulated by 
vacuum polarization \cite{bib4}. 
These contributions are present also in the second equation of the BBGKY chain 
for the correlation functions on the r.h.s. of Eq.~(\ref{1e}) 
(they were omitted in \cite{bib6})
\begin{eqnarray}\label{2e}
\biggl\{\frac{\partial}{\partial t}
&+&i[\omega(\ve{p}_1,t)+\omega(\ve{p}_2,t)-k]\biggr\}
\langle b_{\alpha}(-\ve{p}_{1},t)a_{\beta}(\ve{p}_2 ,t)
A_{r} ^{(+)} (\ve{k},t)\rangle
\nonumber\\
&=&-ie(2\pi)^{-3/2}\int d^3 p' \frac{d^3 k'}{\sqrt{2k'}}\Bigl\{ 
\delta(\ve{p}' -\ve{p}_1 +\ve{k}') 
\nonumber  \\
&& \times \left[[\bar{u}v]^{r'}_{\alpha\beta'}(\ve{p}' ,\ve{p}_1 ,\ve{k}';t)
\langle a^{+}_{\beta'}(\ve{p}',t)a_{\beta}(\ve{p}_2 ,t)
A_{r'}  (\ve{k}',t)A_{r} ^{(+)} (\ve{k},t)\rangle \right. 
\nonumber\\
 &+&\left. [\bar{v}v]^{r'}_{\alpha\beta'}(\ve{p}' ,\ve{p}_1 ,\ve{k}';t)
\langle b_{\beta'}(-\ve{p}',t)a_{\beta}(\ve{p}_2 ,t)
A_{r'}  (\ve{k}',t)A_{r} ^{(+)} (\ve{k},t)\rangle \right]  
\nonumber \\
 &-&\delta(\ve{p}_2 -\ve{p}' +\ve{k}')
 \left[[\bar{u}u]^{r'}_{\beta'\beta}(\ve{p}_2 ,\ve{p}' ,\ve{k}';t)
\langle b_{\alpha}(-\ve{p}_1,t)a_{\beta'}(\ve{p}' ,t)
A_{r'}  (\ve{k}',t)A_{r} ^{(+)} (\ve{k},t)\rangle \right.
\nonumber \\
&+& \left.[\bar{u}v]^{r'}_{\beta'\beta}(\ve{p}_2 ,\ve{p}' ,\ve{k}';t)
\langle b_{\alpha}(-\ve{p}_1,t)b_{\beta'}^+ (-\ve{p}' ,t)
A_{r'}  (\ve{k}',t)A_{r} ^{(+)} (\ve{k},t)\rangle \right]\Bigr\}  
\nonumber \\
 &+&V^r_{\alpha\beta}(\ve{p}_1 ,\ve{p}_2 ,\ve{k};t)~, 
\end{eqnarray}
where the vacuum polarization contributions are collected in the 
following group of terms
\begin{eqnarray}\label{3e}
V^r_{\alpha\beta}(\ve{p}_1 ,\ve{p}_2 ,\ve{k};t)&=&
-ie(2\pi)^{-3/2}\frac{1}{\sqrt{2k}}\int d^3 p_1'd^3 p'_2  
\delta(\ve{p}'_1 -\ve{p}'_2 -\ve{k}) \nonumber  \\
&&\times \left\{[\bar{u}u]^{r}_{\alpha'\beta'}(\ve{p}'_1 ,\ve{p}'_2 ,\ve{k};t)
\langle b_{\alpha}(-\ve{p}_1,t)a_{\beta}(\ve{p}_2 ,t)a^+_{\alpha'}(\ve{p}'_1,t)
a_{\beta'}(\ve{p}'_2 ,t)\rangle  \right.\nonumber \\
&+& [\bar{u}v]^{r}_{\alpha'\beta'}(\ve{p}'_1 ,\ve{p}'_2 ,\ve{k};t)
\langle b_{\alpha}(-\ve{p}_1,t)a_{\beta}(\ve{p}_2 ,t)a^+_{\alpha'}(\ve{p}'_1,t)
b_{\beta'}^+ (-\ve{p}'_2 ,t)\rangle  \nonumber \\
&+& [\bar{v}u]^{r}_{\alpha'\beta'}(\ve{p}'_1 ,\ve{p}'_2 ,\ve{k};t)
\langle b_{\alpha}(-\ve{p}_1,t)a_{\beta}(\ve{p}_2 ,t)b_{\alpha'}(-\ve{p}'_1,t)
a_{\beta'}(\ve{p}'_2 ,t)\rangle  \nonumber \\
&+& \left.[\bar{v}v]^{r}_{\alpha'\beta'}(\ve{p}'_1 ,\ve{p}'_2 ,\ve{k};t)
\langle b_{\alpha}(-\ve{p}_1,t)a_{\beta}(\ve{p}_2 ,t)b_{\alpha'}(-\ve{p}'_1,t)
b_{\beta'}^+ (-\ve{p}'_2 ,t)\rangle  \right\}.
\end{eqnarray}
Here the terms containing anomalous correlators of the type 
$\langle abA^{(+)}\rangle $ 
have been omitted since these correlators vanish in RPA if 
$\langle A^{(\pm)}\rangle =0$.

In order to close the system of Eqs. (\ref{1e}) and (\ref{2e}), let us apply 
now the RPA to the correlators on r.h.s. of these equations, e.g.,
\begin{eqnarray}\label{4e}
   &\langle a^{+}_{\beta'}(\ve{p}',t)a_{\beta}(\ve{p}_2 ,t)
A_{r'}(\ve{k}',t)A_{r}^{(+)}(\ve{k},t)\rangle \simeq
   \langle a^{+}_{\beta'}(\ve{p}',t)a_{\beta}(\ve{p}_2 ,t)\rangle 
\langle A_{r'}(\ve{k}',t)A_{r}^{(+)} (\ve{k},t)\rangle ~.
\end{eqnarray}

The next approximation is the diagonalization of all one-particle correlation 
functions with respect to the momentum variables and spin (or polarization) 
indices,
\begin{eqnarray}
\label{5e}
 \langle A^{(+)}_{r}(\ve{k},t)A^{(-)}_{r'}(\ve{k}',t)\rangle &=&
\delta_{rr'}\delta(\ve{k}-\ve{k'})F_r (\ve{k},t)~,\\
  \langle a^{+}_{\alpha}(\ve{p},t)a_{\beta}(\ve{p}',t)\rangle &=&
\delta_{\alpha\beta}\delta(\ve{p}-\ve{p'})f(\ve{p},t)~,
\label{6e}
\end{eqnarray}
where $f=\frac{1}{2}{\rm tr}_{\rm spin}{f}$ and $F_r(\ve{k},t)$ is the photon 
distribution function with the polarization $r$. 
The relation (\ref{6e}) means that spin effects are neglected.

The approximations (\ref{4e})-(\ref{6e}) allow to rewrite the anomalous 
correlation functions from the l.h.s. of Eq.~(\ref{2e}) taking into account 
the vacuum polar
ization contribution (\ref{3e}) so that
\begin{eqnarray}
\label{7e}
\langle b_{\alpha}(\ve{-p_1},t)a_{\beta}(\ve{p_2},t)A^{(+)}_{r}(\ve{k},t)
\rangle &=& 
-\frac{ie\delta(\ve{p_2}-\ve{p'_1}+\ve{k})}{\sqrt{2k}(2\pi)^{3/2}}\int^t dt'
[\ve{u}\upsilon]^r_{\alpha\beta}(\ve{p_2},\ve{p_1},\ve{k};t')
\nonumber \\
&&\times
\big\{[ f(\ve{p_1},t')+f(\ve{p_2},t')-1][ 1+F_r(\ve{k},t')] \nonumber\\
&&+[ 1-f(\ve{p_1},t')][ 1-f(\ve{p_2},t')]\big\}
{\rm e}^{-i\theta(\ve{p_1},\ve{p_2},\ve{k};t',t)}~,
\end{eqnarray}
where it was used that $f^c=1-f$ due to the electric charge neutrality of the 
vacuum at $t\rightarrow-\infty$ and
\begin{eqnarray}
\label{8e}
\theta(\ve{p_1},\ve{p_2},\ve{k};t',t)
=\int^t_{t'}d\tau \left[\omega(\ve{p_1},\tau)+\omega(\ve{p_2},\tau)-k\right]~.
\end{eqnarray}
The first group of terms in the curly brackets in Eq.~(\ref{7e}) corresponds 
to the one-photon annihilation process (this contribution was investigated in
the works \cite{bib6}) while the second group describes the radiationless 
vacuum fluctuations. 
In the case of a strong subcritical "laser" field the number density of the 
radiated photons is not large, $F_r(\ve{k},t)\ll 1$, so that the influence of 
the photon reservoir on the photon emissivity of the system can be neglected. 
Eq.~(\ref{7e}) then takes the form
\begin{eqnarray}
\label{9e}
\langle b_{\alpha}(\ve{-p_1},t)a_{\beta}(\ve{p_2},t)A^{(+)}_{r}(\ve{k},t)
\rangle 
&=&-\frac{ie\delta(\ve{p_2}-\ve{p'_1}+\ve{k})}{\sqrt{2k}(2\pi)^{3/2}} 
\int^t dt'{\rm e}^{-i\theta(\ve{p_1},\ve{p_2},\ve{k};t',t)}
\nonumber\\
&&\times [\ve{u}\upsilon]^r_{\alpha\beta}(\ve{p_2},\ve{p_1},\ve{k};t')
f(\ve{p_1},t')f(\ve{p_2},t')~.
\end{eqnarray}
Substituting (\ref{9e}) into Eq.~(\ref{1e}), we obtain a closed expression for 
the photon production rate
\begin{eqnarray}
\label{10e}
 \dot{F}(\ve{k},t) =\frac{e^2}{4k(2\pi)^{3}} \int^{t}dt'\int d^3 p 
{\rm e}^{-i\theta(\ve{p},\ve{p}+\ve{k},\ve{k};t',t)}
K(\ve{p},\ve{p} +\ve{k},\ve{k};t,t') 
f(\ve{p} ,t')f(\ve{p}+\ve{k},t') + c.c.,
\end{eqnarray}
where we have introduced the two-time convolution
 \begin{eqnarray}\label{11e}
 K(\ve{p} ,\ve{p}+\ve{k},\ve{k};t,t')
=[\bar{v}u]^{r}_{\beta\alpha}(\ve{p} ,\ve{p}+\ve{k} ,\ve{k};t)
~[\bar{u}\upsilon]^{r}_{\alpha\beta}(\ve{p}+\ve{k},\ve{p},\ve{k};t').
\end{eqnarray}
Additionally, it is assumed in Eq.~(\ref{10e}) that the photons have 
equiprobable distributions regarding their polarizations, $F_1=F_2=F$.

Thus, the photon production rate is a nonlinear (quadratic) non-Markovian 
function with respect to the electron-positron distribution function 
$f(\ve{p},t)$. 
This effect corresponds to the result \cite{bib8} of the S-matrix approach.

The consequent estimation procedure of the integrals on the l.h.s. of 
Eq.~(\ref{10e}) (method of the photon count) was presented in \cite{bib6}.
The meaning of these approximations is the following. 
On the r.h.s. of Eq.~(\ref{10e}) there is a high frequency multiplier 
$exp\{-i\theta\}$ with the phase (\ref{8e}). 
In order to select the low frequency component of the photon production rate 
(\ref{10e}) (only this corresponds to the observable value), it is necessary 
to compensate this high frequency phase by means of the higher harmonics in 
the Fourier decompositions of the other functions in the integral (\ref{10e}). 
According to the structure of the time dependent $u,\upsilon$-spinors 
\cite{bib4,bib6}, the convolution (\ref{11e}) is a polynomial in $e A(t)$ 
(below it is assumed that the "laser" electric field is 
$A(t)=A^{3}(t)=-(E_0/\nu) \cos(\nu t)$) and hence it can not garantee for the 
necessary compensation. 
Therefore, we use here the Markovian approximation 
$K(\ve{p},\ve{p}+\ve{k},\ve{k};t,t')\rightarrow 
K(\ve{p},\ve{p}+\ve{k},\ve{k};t,t)\simeq K_0\approx 5$. 
For large adiabaticity parameters $\gamma=E_c\nu/Em \gg1$, we obtain then in 
the low frequency approximation ($\alpha=e^2/4\pi$)
\begin{eqnarray}
\label{12e}
\dot{F}(k)=\frac{\alpha K_0}{2k}
\frac{p_1\omega_1\sqrt{\omega_1^2+k^2}}{\omega_1+\sqrt{\omega_1^2+k^2}}
J_{n_0+1}(a)\left[ J_{n_0+3}(a)+J_{n_0-1}(a)\right]f_2(p_1)f_2(p_1+k)~,
\end{eqnarray}
where we took into account the lowest harmonics of the distribution function, 
$f_2(p)$ (we model the time dependence of the distribution function as 
$f(\ve{p},t)=f(p)[1-\cos 2\nu t]/2$), $\omega_1=\sqrt{m^2+p_1^2}$, and 
$J_n(a)$  is the Bessel function. 
Its argument is
\begin{eqnarray}\label{13e}
a=\frac{2\sqrt{\pi\alpha}E_0}{\nu^2}\left[ \frac{p_1}{\omega_1}
+\frac{p_1+k}{\sqrt{\omega_1^2+k^2}}\right] ,
\end{eqnarray}
With $p_0$ we denote the positive root of the equation $\Omega_0-n\nu=0$, where
\begin{eqnarray}\label{14e}
\Omega_0=\omega_1+\sqrt{\omega_1^2+k^2}-k
\end{eqnarray}
is the mismatch and $n_0=\left[\Omega_0(p_0=0)/\nu \right]$ 
($[x]$ is the integer part $x$) is the photon number necessary for 
overcoming the energy gap. Then we obtain
\begin{eqnarray}\label{15e}
 p_0=\left\{\frac{(n\nu)^2(n\nu-2k)^2}{4(n\nu -k)^2}-m^2\right\}^{1/2}~.
\end{eqnarray}

\subsection{The case of optical vacuum excitation ($\nu\ll m$)}

In the optical part of the photon spectrum ($k \lesssim \nu$) we have 
$p_1$=$\sqrt{km}$ as the first root of the equation 
$\Omega_0(p_1)-(n_0+1)\nu=0$ (it corresponds to the leading contribution from 
the set $n>n_0$), $a=2(E_0/E_c)(m/\nu)^{3/2}\ll 1$ and $n_0=[2m/\nu]\gg 1$, 
i.e., the necessary number of quasiclassical photons is huge. 
However, it defines a very small intensity of photon radiation. 
From Eq.~(\ref{12e}) it follows that
\begin{eqnarray}\label{16e}
I(k)=dF(k)/mdt
=\frac{1}{4}\alpha K_0\sqrt{\frac{m}{k}}f_2^2(0)
\left\lbrace \frac{1}{n_0^2}\left[
\frac{E_0}{E_c}\left(\frac{m}{\nu}\right)^{3/2}\right]^{2n_0}\right\rbrace~.
\end{eqnarray}
The value $f_2(0)$ can be estimated as a result of the numerical solution of 
the KE for $e^- e^+$ excitations in a "laser" field. 
For the PW laser system Astra Gemini we have 
$E_0\sim 10^{-5}E_c$, $\lambda=800$ nm and $f_2(0)\sim 10^{-11}$ \cite{bib3}. 
The spectral distribution of the radiated photons from the volume $\lambda^3$ 
of the focal shot per second,
\begin{eqnarray}\label{17e}
\frac{dN_k}{dtdk}=\frac{\lambda^3}{\pi^2}Ik^2=\frac{8\pi k^2 m}{\nu^3}I~,
\end{eqnarray}
will be negligibly small for the mentioned parameters. 
However, the term in the curly brackets on the r.h.s. of Eq.~(\ref{16e}) 
behaves as a $\theta(a)$-function with the branch point $a_0=2$ when 
letting $a\rightarrow a_0$. 
Then the spectral distribution (\ref{17e}) starts to grow strongly. 
Unfortunately, this value $a_0=2$ lies outside of the validity range of
Eq.~(\ref{16e}). nevertheless, this gives a hint on the possible growth of the 
radiation intensity in this domain. Some additional analysis is necessary here.

The $\gamma$-ray part of the photon spectrum $k \sim m$ can not be considered 
in the framework of this approximation ($a\gg 1$ again).

\subsection{The case of $\gamma$-ray vacuum excitation ($\nu\sim m$)}

In this case the mismatch (\ref{14e}) can be compensated by the smallness of 
the photon number from the quasiclassical "laser" field, $n_0\gtrsim 1$. 
Let $n_0=1$ (this is the hypothetical limiting case; for the planned XFEL 
facility with $\lambda=0.15$ nm \cite{bib9}).
The developed theory is working well in this case ($a\ll 1 $) for the 
subcritical fields $E\ll E_c$.

In the optical part of the photon band ($k\ll m$) we have $p_1=\sqrt{3}m$ and
\begin{eqnarray}\label{18e}
a=\sqrt{3}\frac{E_0}{E_c}\left(\frac{m}{\nu} \right)^2~.
\end{eqnarray}
The spectral distribution following Eq.~(\ref{12e}) is
\begin{eqnarray}\label{19e}
\frac{dN_k}{dtdk}=\frac{3\sqrt{3}\pi \alpha K_0 k}{2 \nu}f_2^2(p_1)
\left(\frac{E_0}{E_c} \right)^2\left(\frac{m}{\nu} \right)^6.
\end{eqnarray}
Thus, the effect grows linearly with $k$. 
For $\nu=1$ MeV, $E_0=10^{-5}E_c$ and $f_2(p_1)\sim 10^{-11}$ we obtain again 
a negligibly small effect: the suppression factor is 
$f_2(0)E_0/E_c \sim 10^{-5}$ so that a very weak signal results.

The situation is slowly changing when going to higher frequencies of the
excited signal ($X$-ray or $\gamma$-ray domain)  at $E/E_c=$ const. 
One can demonstrate this by writing $p_1$ (\ref{15e}) for $k\neq 0$ and $n_0=1$
\begin{eqnarray}\label{20e}
p_1=m\left\lbrace 4\left[ \frac{2+k/m}{2+k/2m}\right]^2-1\right\rbrace^{1/2}~.
\end{eqnarray}

The situation becomes more optimistic at $E\rightarrow E_c$ when $a\to 1$. 
For example, for the XFEL with $E=0.24~E_c$ and $\lambda=0.15$ nm \cite{bib9} 
the intensity (\ref{19e}) can be accessible to observation, apparently. 
However, this case needs special investigation since for $\gamma\lesssim 1$ 
the presented approach is not valid.

\section{Summary}

For the probability estimation of the photon radiation from the focal spot of 
counter propagating laser beams we have shown that the one-photon annihilation 
channel in the domain of the large adiabaticity parameter
$\gamma=E_c\nu/(Em)\gg 1$ does not lead to an appreciable effect. 
However, decreasing $\gamma\rightarrow 1$ shows the tendency to increase the 
photon radiation intensity  and thus the possibility to observation secondary 
photons  for some domains of the laser field parameters as, e.g., for the XFEL.
Just in this domain there is a significance for experimental observability 
which warrants further theoretical investigation of this case.

\subsection*{Acknowledgements}
We thank G. Gregori, B. K\"ampfer, C.D. Murphy and S.M. Schmidt for their 
encouragement and interest in the progress of this work. 
S.A.S. is grateful for support of the FZ J\"ulich for his visit and for the 
hospitality extended to him.  
D.B. acknowledges support from the Russian Fund for Basic Research, grant
No. 08-02-01003-a and from the Polish Ministry for Science and Higher 
Education, grant No. NN 202 231837.

\end{document}